\documentclass[conference]{IEEEtran}
\IEEEoverridecommandlockouts
\usepackage{cite}
\usepackage{amsmath,amssymb,amsfonts}
\usepackage{algorithm}
\usepackage{algorithmic}
\usepackage{graphicx}
\usepackage{textcomp}
\usepackage{xcolor}
\def\BibTeX{{\rm B\kern-.05em{\sc i\kern-.025em b}\kern-.08em
		T\kern-.1667em\lower.7ex\hbox{E}\kern-.125emX}}
\begin{document}
	
	\title{Mixing detection on Bitcoin transactions using statistical patterns
	}
	
	\author{\IEEEauthorblockN{1\textsuperscript{st} Ardeshir Shojaeenasab}
		\IEEEauthorblockA{\textit{School of Electrical and}\\ \textit{Computer Engineering} \\
			\textit{University of Tehran}\\
			Tehran, Iran \\
			a.shojaeenasab@ut.ac.ir}
		\and
		\IEEEauthorblockN{2\textsuperscript{nd} Amir Pasha Motamed}
		\IEEEauthorblockA{\textit{School of Electrical and}\\ \textit{Computer Engineering} \\
			\textit{University of Tehran}\\
			Tehran, Iran \\
			a.motamed@ut.ac.ir}
		\and
		\IEEEauthorblockN{3\textsuperscript{rd} Behnam Bahrak}
		\IEEEauthorblockA{\textit{School of Electrical and}\\ \textit{Computer Engineering} \\
			\textit{University of Tehran}\\
			Tehran, Iran \\
			bahrak@ut.ac.ir}
	}
	
	\newtheorem{definition}{Definition}
	
	\maketitle
	
	\begin{abstract}
		Cryptocurrencies gained lots of attention mainly because of the anonymous way of online payment, which they suggested. Meanwhile, Bitcoin and other major cryptocurrencies have experienced severe deanonymization attacks. To address these attacks, Bitcoin contributors introduced services called mixers or tumblers. Mixing or laundry services aim to return anonymity back to the network. In this research, we tackle the problem of losing the footprint of money in Bitcoin and other cryptocurrencies networks caused by the usage of mixing services. We devise methods to track transactions and addresses of these services and the addresses of dirty and cleaned money. Because of the lack of labeled data, we had to transact with these services and prepare labeled data. Using this data, we found reliable patterns and developed an integrated algorithm to detect mixing transactions, mixing addresses, sender addresses, and receiver addresses in the Bitcoin network. 
	\end{abstract}
	
	
	\section {Introduction}
	Along with the growing popularity of cryptocurrencies and their usage in different contexts, financial crimes have also increased significantly in these services. One of the most deceiving tools that bury the link between sender and recipient of money in cryptocurrencies is money laundering or mixing services. Combining different sources of money with dirty money makes tracking dirty money complex and obscure and distorts the relationship between the sender and receiver of money. Due to this feature, money laundering services are widely used to eliminate the track of income from ransomware, thefts, sales of weapons and drugs, and other illegal activities by combining these incomes with other sources of money and disrupting the tracking of these incomes.

Finding a relationship between sender and recipient of a mixing process is an attack against the anonymity caused by mixing services in Bitcoin and other cryptocurrencies. This deanonymization attack comprises four steps that are dependent on each other. First, we should find transactions corresponding to the mixing services through the blockchain. Second, we should classify the filtered transactions into incoming, intermediate, and withdrawing transactions. Third, we should label addresses in the classified transactions of the second step using three types of labels: (1) sender addresses, (2) mixing service addresses, and (3) receiver addresses. Finally, using these labeled addresses, we should try to link sender and receiver addresses that correspond to each other in a real-world transaction and eliminate the anonymity caused by the mixing services.
	
	One of the main challenges we encountered in this study was the lack of labeled data for mixing services transactions. Due to the anonymous nature of the mentioned problem, nobody shares their mixing transactions due to anonymity loss and privacy concerns. It means we cannot perform many standard methods for pattern recognition that depend on a large amount of labeled data. 

In this paper, we propose a method to achieve following goals:
\begin{itemize}
\item
Identify the transactions of three well-known mixing services based on their functional characteristics.
\item
Classify those transactions and their addresses into the input, hidden internal, and output transactions.
\end{itemize}
	
	\section{Related Work}
	On the 21\textsuperscript{st} of January 2009, Hal Finney tweeted; "Looking at ways to add more anonymity to bitcoin" \cite{haltweet}. This tweet was published less than two months after publishing the Bitcoin whitepaper \cite{Satoshi} and indicates that the anonymity of Bitcoin from the beginning of this technology was a serious concern of its pioneers.
	
	During the Bitcoin lifetime, many researchers worked on the anonymity of Bitcoin both in breaking the anonymity of the network and helping to add more anonymity levels to the network. In some of these studies, researchers tried to beat the anonymity of Bitcoin at the network level. In \cite{kaminsky}, Kaminsky suggested that the first node in the p2p network of Bitcoin that delivers a transaction is most likely the source of the transaction. In subsequent attempts to use this idea for deanonymization, Koshy et al. studied transaction information at the network layer, deciding whether the received transaction was relay or non-relay, who owned the transaction, and the relevant IP address \cite{Koshy2014p2ptraffic}. Biryukov et al., in a 2014 study, were able to come up with ways even to cross the firewall and network address translator (NAT) at the network level \cite{Bir2014p2pnetwork}. They also categorized the obtained IP addresses and the transaction ratios for each IP, bringing the clustering of addresses to another level \cite{Bir2019NetworkAnal}.
	
	Another type of research on the anonymity of Bitcoin focuses on investigating the transaction's structures.
	One of the most effective attacks on Bitcoin's anonymity is an attack on related addresses. This attack allows the addresses associated with a single identity to be grouped into clusters. It means that if one of these addresses is later linked to a real-world identity, the identity of the other addresses on the same cluster can be recognized. One of the significant studies with this approach was the research of Mikeljan et al. on the Silk Road store at its peak of activity \cite{MaeiklejonFistful}. They suggested the idea that in order to find the addresses of different services, we should transact with them. After each transaction, they used the obtained addresses to expand the cluster of addresses by experimental methods such as assuming that the real-world identity of all addresses contributing to a transaction's input is the same.
	In \cite{PatternIdentification}, Chang et al. retrieved repetitive patterns at the transaction level in the Bitcoin blockchain and were able to find patterns such as relays, sweepers, and distributors.
	
	More related to this paper are studies focused on address labeling tasks and mixing detection. There are some methods for adding more anonymity to Bitcoin, such as CoinJoin \cite{CoinJoin}, generating new addresses, mixing services and stealth addresses \cite{stealth}. For more, de Moser et al. published a comprehensive survey on methods of adding more anonymity to Bitcoin  \cite{Moser2017AnonyAlone}.
	
	In \cite{Moser2013}, the authors demonstrated attacks on three different mixing services that were active in 2013 and targeted their mixing methods. In one of the services, they succeeded in finding a mixing method using taint analysis and found that in some cases, the footpath between sender and receiver was not completely cleared, and there was a connection between them. Today's mixing services have become much more intelligent, and we cannot attack them using taint analysis due to disconnection between sender and receiver.
In a white paper published in 2015 by the network security company Novetta \cite{Novetta},  the authors targeted three mixing services from a structural point of view and transacted with them. They concluded that taint analysis is not helpful anymore when mixers are used. Moreover, they made another conclusion that although analyses and patterns can be found for mixing services, a direct connection between sender and receiver could not be established, and mixers are effective tools for hiding true identity in the Bitcoin network. Another study by Baltazar and Castro of Chainalysis examined three mixing services and tried to find some patterns \cite{balthazar}. This work showed that mixing services follow a specific pattern in their activity, but they did not release any specific pattern. 
		
	Sun and Yang proposed a method for mixing detection using long short-term memory (LSTM) and claimed they achieved a high recall approach in contrast to previous pattern-based methods \cite{LSTM}. The main disadvantage of their work is using an old dataset of the Bitcoin network belonging to 2013.

	In a recent study, Pakki et al. did a comprehensive research on several common mixing services from different aspects \cite{pakki}. This study is helpful for understanding mixing services operation, but this study does not consider the functional behavior of mixers in terms of transactions and addresses it creates in the blockchain. In this paper, we focus on the patterns of mixing services in terms of their transactions and addresses.
	
	\section{Methodology}
	This section will explain our methodology, including collecting required data and extracting common patterns in the mixing transactions, eventually leading to an algorithm for filtering such transactions.

	\subsection{Data Gathering}\label{AA}
	To investigate mixing transactions, we need labels for transactions that participated in mixing services processes, and we need access to the full blockchain transactions information.
	\subsubsection{Retrieving Blockchain Data}
	To access transactions data within the blockchain, we used a full node Bitcoin client and collected the data we needed from it using its JSON-RPC APIs. After running the Bitcoin client, according to its implemented RPC functions, information about blocks and transactions, such as block hash, transactions input and output addresses, transferred values, timestamps, and other required information included in the blockchain, can be retrieved.

	\subsubsection{Collecting Labels for Mixing Transactions}
	Mixing services are used by individuals with different intentions, for example, to conceal their financial transactions for tax evasion, criminal purposes, or privacy concerns. What is clear is that no matter what their intentions are, individuals are reluctant to disclose their mixing transactions, and as expected, there is no dataset of transactions related to mixing services. Although some labels for money laundering service addresses are available, errors may occur due to clustering methods and heuristics that label other addresses according to the previously labeled addresses. During this process, ordinary addresses may be attributed to money laundering services, and studying their transactions will be misleading. Due to the lack of an intended dataset, we needed to collect accurate data independently.
	
	To collect mixing transactions, we selected three of the most well-known mixing services that were primarily discussed in BitcoinTalk\footnote{https://bitcointalk.org} and Reddit\footnote{https://www.reddit.com} to collect a dataset of mixing transactions. Using their service and depositing money to their specified address and returning money through the laundering process from them to our recipient address, we made two mixing transactions (one for deposit and the other for withdrawal) each time we used these services. Table~\ref{tab:mixingservicesTried} lists these three selected mixing services and the number of mixing transactions we created for each.
	
	\begin{table}[ht]
		\caption{Mixing services that we gathered data from. }
		\label{tab:mixingservicesTried}
		\centering
		\begin{tabular}{|c|c|c|}
			\hline
			\textbf{Name}	&  \textbf{URL} & \textbf{Number of Transactions} \\
			\hline
			MixTum	& mixtum.io	& 12 \\
			\hline
			Blender & blender.io	&  4 \\
			\hline
			CryptoMixer	& cryptomixer.io	&  6 \\
			\hline
		\end{tabular}
	\end{table}
	
	\subsection{Extracting Statistical Pattern}
	Due to the financial and time constraints of collecting a rich dataset of mixing transactions, we had to extract the mixing pattern using simple statistical methods that are suitable for small data. 
	
	First, we discuss the mechanisms used by different mixing services briefly. We describe two main types of mixing services. 
	
	\begin{figure}
		\centering
		\includegraphics[width=0.95\linewidth]{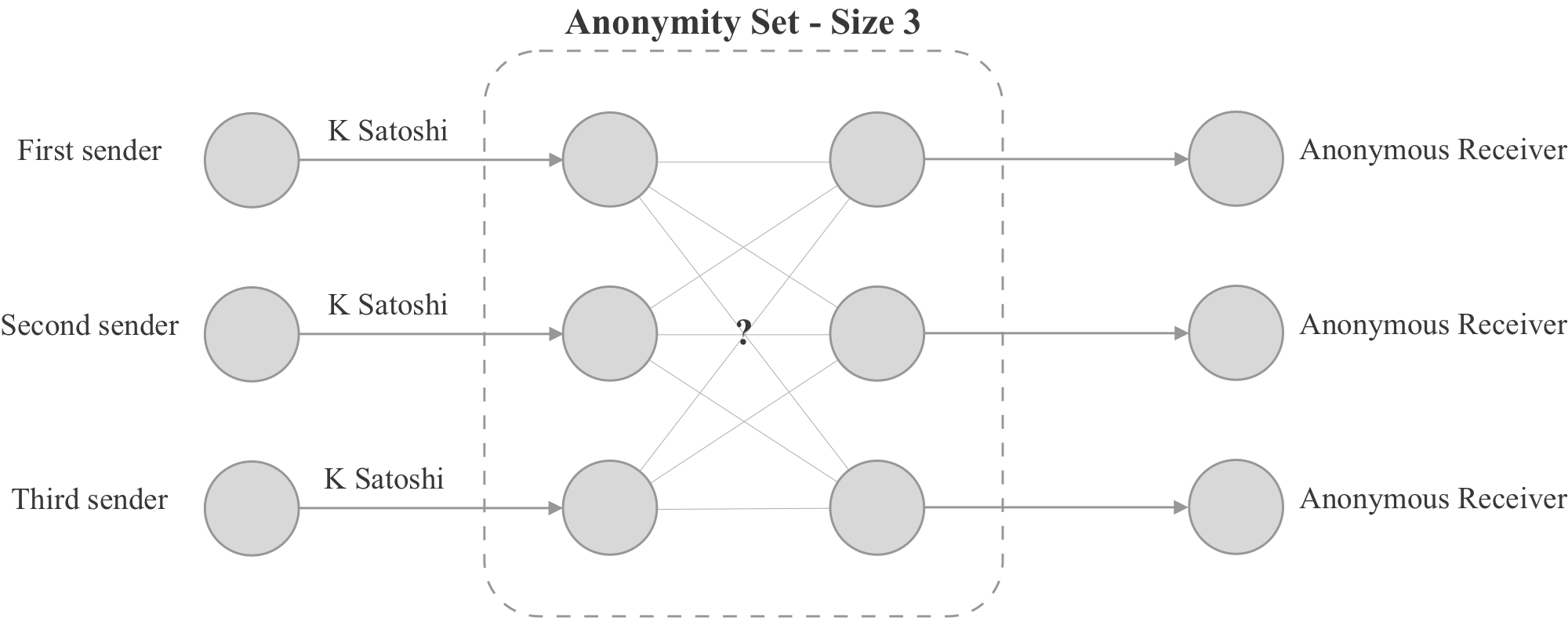}
		\caption{The general structure of traditional mixing services.}
		\label{fig:tradMixing}
	\end{figure}
	
\begin{figure}
		\centering
		\includegraphics[width=0.95\linewidth]{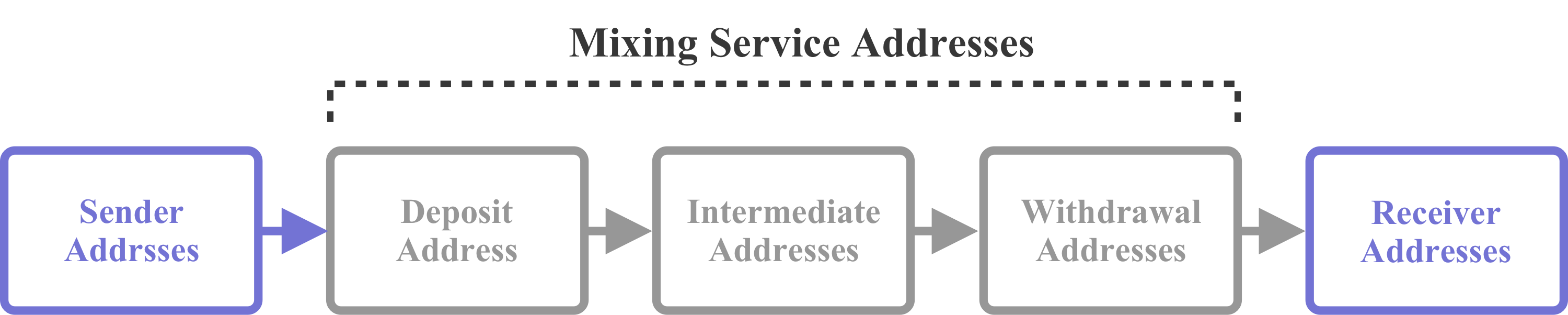}
		\caption{Mixing service structure in general}
		\label{fig:modernMixing}
	\end{figure}

	\begin{itemize}
		\item \textbf{Traditional mixing services}: In traditional mixing services, the service determines a fixed amount of money that can be mixed with the money from requested transactions for mixing. These services exchange money from senders to receivers randomly. In the first step, the service creates a new random deposit address for the sender. Each time the user refreshes the mixing request's web page, the address is regenerated. In the next step, a sender needs to wait until at least $N-1$ other mixing requests with the same amount are registered and deposit their money into their generated deposit address. The mixing service in this stage, which is called the intermediate stage, creates transactions in which these $N$ inputs are interchanged randomly and transferred to $N$ anonymous output addresses to clear the footprint of money transfer between the sender and the recipient. This operation sets the anonymity set to a probability of $1/N$, which shows the probability of finding a correct link between a sender and its corresponding recipient. Figure \ref{fig:tradMixing} illustrates how these services work.
		
		\item \textbf{Modern mixing services}: These services, like traditional services, consist of three phases. First, the requested money is sent to a random deposit address created by the mixing service. The main difference is here; The mixer does not need to wait to receive money from other individuals and delivers cleaned money to the recipient from one of its addresses where it has sufficient money. 
This way, the footprint is wholly erased because the receiving and sending transactions in the blockchain are not related. Also, due to the specified delay, tracking money becomes more difficult. It should be noted that this method has no limit in terms of the amount of money for laundering. In figure \ref{fig:modernMixing}, we put the abstract mechanism of this type of mixing service. In this paper, we aim to attack modern mixing services.
	\end{itemize}
	
	Although the traditional method of money laundering in Bitcoin added some anonymity and cleared the path of money transfer, it had serious weaknesses that led to the emergence of modern mixing services. The most severe limitation of the traditional method was poor service time which individuals had to wait for another $N-1$ mates to mix their money. To be more anonymous in the network, individuals had to mix their money with a sufficiently large number of mates. However, since the number $N$ corresponds to mix mates and cannot be high, the transferred money path was not completely cleared, and there is a good chance that the transaction will be tracked. In other words, in this method anonymity level is correlated directly with the value of $N$ (anonymity parameter).
	
	In contrast, some advanced parameters make modern mixing services more intelligent. The first parameter is a random time delay. Today most services offer a random time of 1 to 1440 minutes between receiving and sending transactions, making it more difficult to track the money footpath based on time patterns. The second parameter is the limit for the number of recipient addresses which most of the time in modern mixing services is infinite. In fact, in these services, the recipient addresses can be considered as much as desired, making money tracking very difficult due to departed money portions in the destination.
	
	When there is no money directly sent in the transfer cycle to the destination addresses, tracking and finding a link between the first stage of money laundering part to the second and third stages is practically failed, and determining this path is very challenging.
	
	\subsection{First Phase of Detection; Transaction-Level Patterns}
	By investigating mixing transactions we created, we saw recurring patterns. These patterns were found in withdrawal transactions. At the transaction level, we could not find any pattern in the deposit transactions because the structure of deposit transactions was typical in the whole blockchain, and senders created them by using their wallet application to deposit money to the mixing service. Nevertheless, the withdrawal transactions created by the mixing service were discriminative and followed a specific pattern that we will explain.
	
\begin{figure}
		\centering
		\includegraphics[width=0.95\linewidth]{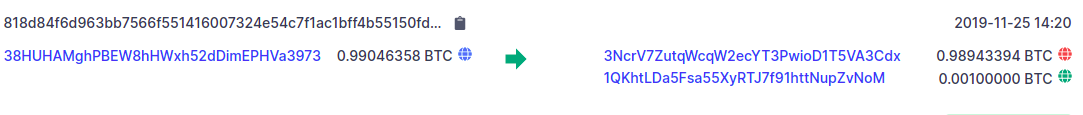}
		\caption{Sample withdrawal transaction of MixTum Service}
		\label{fig:withdrawalTx}
	\end{figure}

	In figure \ref{fig:withdrawalTx}, we show one of the transactions received from the Mixtum mixing service due to using its service.
	
	\subsubsection{Inputs and Output Counts Pattern}
	We found a series of common patterns that can be turned into detection algorithms by investigating the transactions. In all transactions received from mentioned mixing services, the withdrawal transactions were $1:2$. In other words, they had one input and two outputs, one of which is the output address of the money laundering destination (recipient of clean money), and the other for the change address of the mixing service, which deposits the rest of the remaining money at a new address.
	
	\begin{figure}
		\centering
		\includegraphics[width=0.9\linewidth]{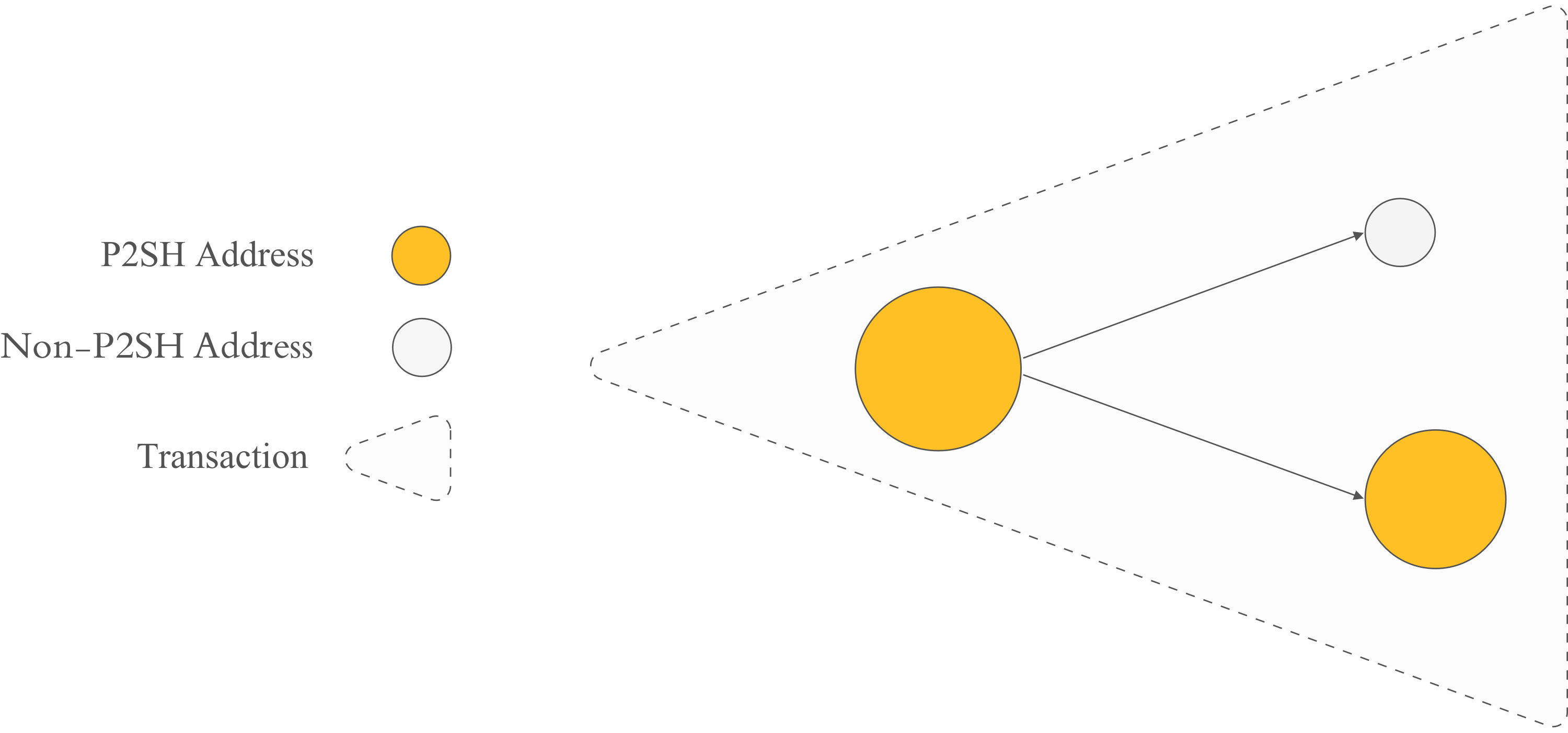}
		\caption{Pattern of suspected mixing withdrawal transaction in first phase}
		\label{fig:suspectedTx}
	\end{figure}
	
	\subsubsection{Address Types Pattern}
	By investigating the type of addresses within the 1:2 structure of candidate transactions that we noticed earlier, we inferred the pattern that has two main characteristics based on withdrawal transactions of Blender and MixTum services:
	\begin{itemize}
		\item Input address type is P2SH.
		\item At least the type of one of the two output addresses is P2SH.
	\end{itemize}
	It is noteworthy that the above patterns were not fully matched to the CryptoMixer service.
	
	\subsubsection{Pattern of Amount Fraction in Transfer}
	Another pattern observed in the withdrawal transactions of mixing services was the patterns of the amount fraction of each output. The input in this 1:2 transaction has a high value of Bitcoin, and a threshold of more than 1 Bitcoin can be considered as an appropriate value for it. Similarly, the amount of output, which is P2SH, in 97\% of cases, is more than five times greater than the amount of the other output (non-P2SH output).
	
	In figure \ref{fig:suspectedTx} we show the structure and characteristics of mixing withdrawal transactions in the schematic, which is used in the first phase filter of transactions.
	
	\subsection{Second Phase of Detection; Chain-Level Patterns}
	After applying the above filter at the transaction level and checking the output, we realize that the detected transactions have many errors. In general, many exchanges and wallets follow similar patterns. This similarity of patterns causes these transactions to be mistakenly identified as mixing transactions in the first phase. Due to the high false-positive rate of the above pattern, we needed to develop a two-phase filter algorithm to act as a complement to the above pattern.
	
	\subsubsection{Footprint of Deposit Transactions in the Mixing Services}
	So far, we have only investigated the mixing withdrawal transaction, and we did not consider deposit transactions and their position in identifying mixing transactions. Nevertheless, in this section, we want to answer the question of whether the incoming transaction to the mixing service also provides us with discriminative information to increase the accuracy, reduce false positives, and increase recall of the detection method or not.
	
	\subsubsection{Sweeper Transactions}
\begin{figure}
		\centering
		\includegraphics[width=0.6\linewidth]{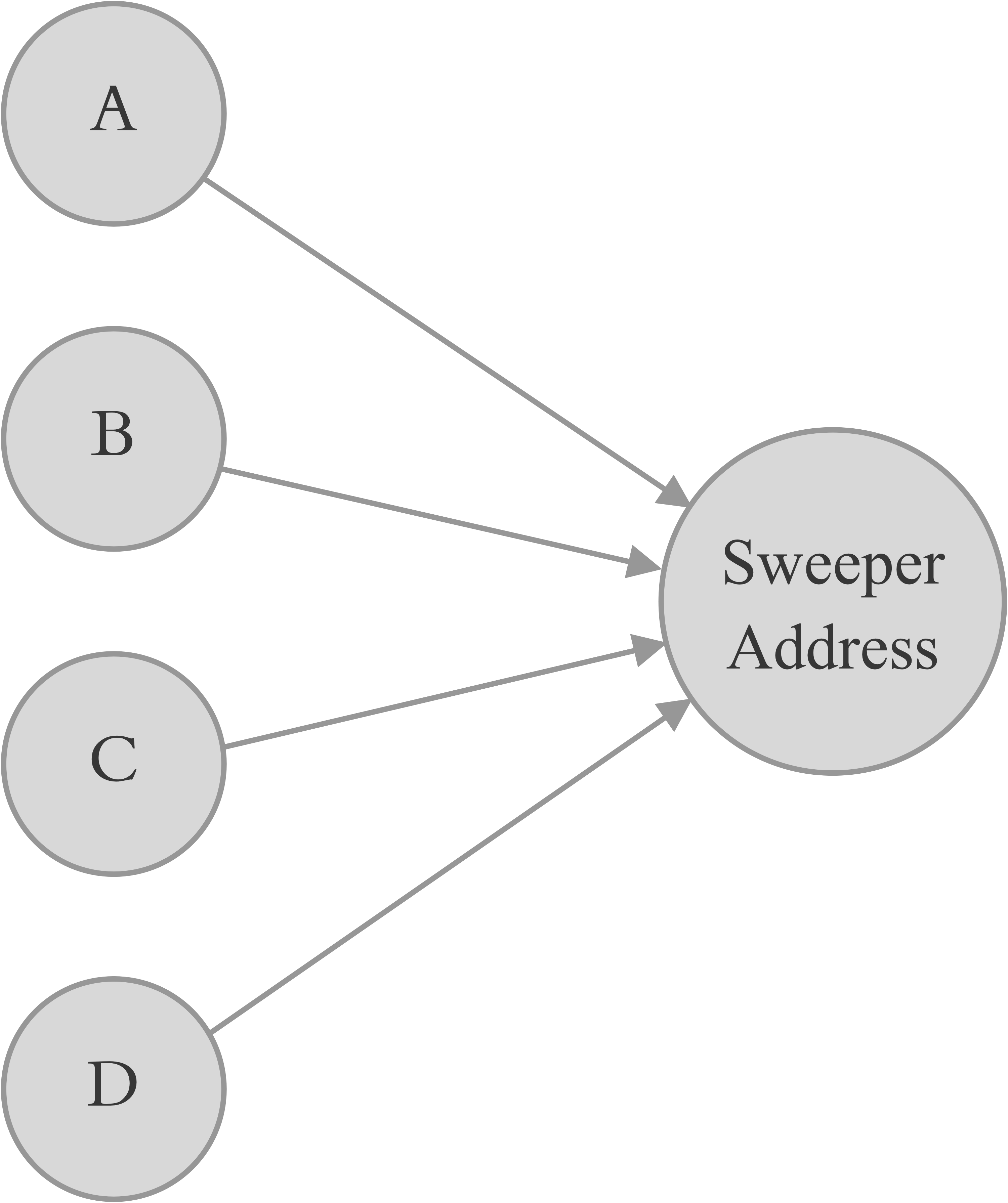}
		\caption{Schematic view of sweeper transaction}
		\label{fig:sweeperTx}
	\end{figure}

	In a study, Chang et al. named several repetitive transaction patterns in the Bitcoin network, one of which they called a sweeper \cite{PatternIdentification}. Sweeper transactions are transactions that, according to the image \ref{fig:sweeperTx}, have a large number of input addresses and one or two output addresses. In other words, transactions that combine different scattered monies deposited in a large number of addresses into one or two addresses, and by executing this transaction, produce a new address with a high amount of money. In other words, the structure of these transactions is often many-to-one.
	
	\subsubsection{Transaction Chain Definition}
\begin{figure}
		\centering
		\includegraphics[width=0.95\linewidth]{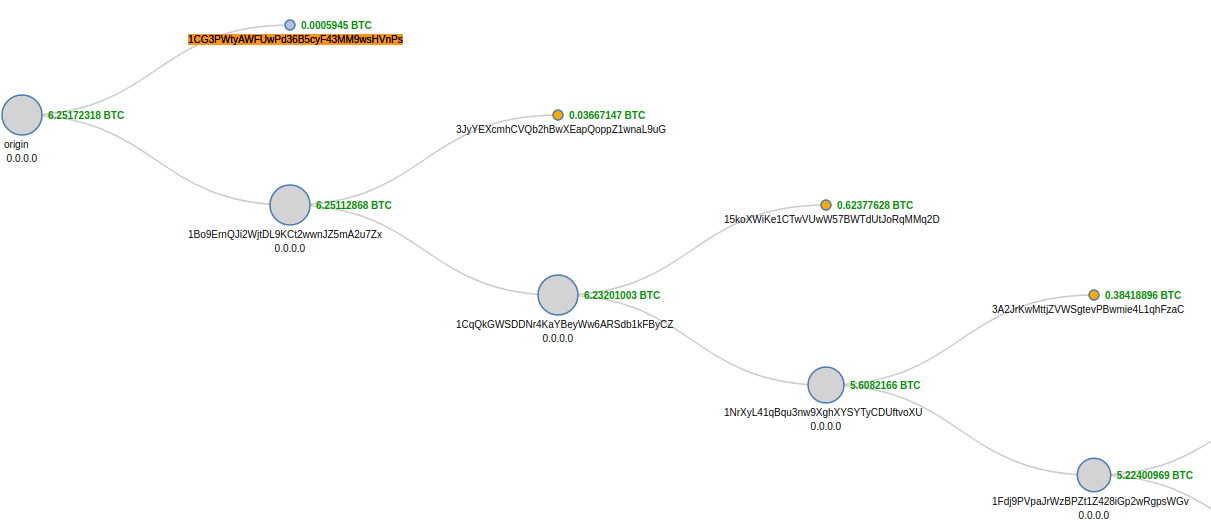}
		\caption{Tree view of transaction chain of an address we got out cleaned money from a mixing service}
		\label{fig:txChainTree}
	\end{figure}

\begin{figure}
		\centering
		\includegraphics[width=0.95\linewidth]{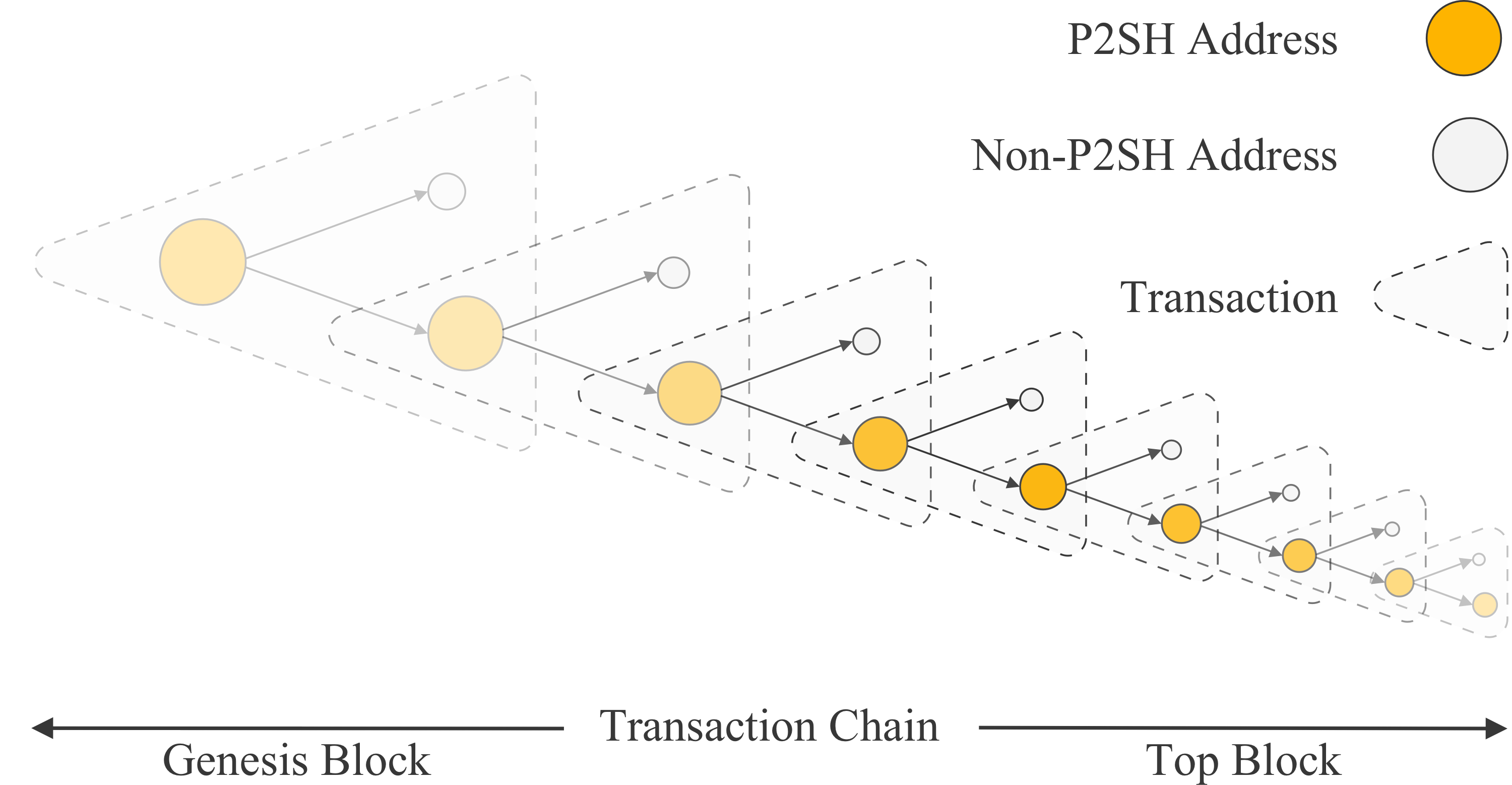}
		\caption{Schematic view of the transaction chain}
		\label{fig:txChainSchematic}
	\end{figure}

	To explain the second phase of the filter, we needed to define a new concept called the transaction chain, which we describe below.
	\begin{definition}
		The sequence $t_1$,$t_2$,$t_3$,..,$t_n$ is a transactions chain where $t_i$ is a transaction and $OUT(t_i)$ is the set of outputs of transactions $t_i$ and $IN(t_i)$ is the set of inputs of transactions $t_i$ and for each $i$ ($0 \leq i \leq n-1$):
		\begin{equation}
			\centering
			\exists a : a \in OUT(t_i) \land a \in IN(t_i+1)
		\end{equation}
	\end{definition}
	A transaction chain is a sequential chain of transactions in which, at each point of the chain, at least one of the outputs is spent in at least one of the inputs of the next transaction. In figure \ref{fig:txChainTree}, as it can be seen, we better visualized the transaction chain as a tree, with each node corresponding to an address on the Bitcoin network, the direction of the edges representing the transfer from source to destination, and the value on each edge indicates the amount of transferred money between addresses in the chain.
	In figure \ref{fig:txChainSchematic} we provided a schematic view of the structure of the mixing transaction chain according to our inference till now. 
	
	\subsubsection{Implementing Algorithm of Identifying Mixing Transaction Chain}
	In this section, we will implement the algorithm for obtaining a transaction chain of mixing service from a candidate mixing transaction.
	
	\begin{algorithm}[H]
		\caption{Detecting transaction chain of mixing service based on candidate transactions} \label{alg2}
		\textbf{Input}:
		A candidate intermediate mixing transaction\\
		\textbf{Output}: Full Chain of Mixing Transactions , Anomalies Detected
		\begin{algorithmic}[1]
			\STATE $\mathbf{AnomalyThreshold} \leftarrow \mathbf{4} $
			\STATE $\mathbf{AnomalyDetected} \leftarrow \mathbf{0} $
			\WHILE{Meet a sweeper transaction, Go Backward} 
			\IF{AnomalyDetected == AnomalyThreshold} \STATE {Break} \ENDIF
			\IF{Algorithm1(transaction) == False} \STATE {AnomalyDetected++} \ENDIF
			\STATE Append transaction hash to the chain list
			\ENDWHILE
			\WHILE{While greater output is spent, Go Forward} 
			\IF{AnomalyDetected == AnomalyThreshold} \STATE {Break} \ENDIF
			\IF{Algorithm1(transaction) == False} \STATE {AnomalyDetected++} \ENDIF
			\STATE Append transaction hash to the chain list
			\ENDWHILE
		\end{algorithmic}
	\end{algorithm}
	
	As can be seen in algorithm \ref{alg2}, we go backward from a feed transaction until meeting a sweeper transaction. Also, we go forward until the end of a transaction chain and when outputs are not spent anymore. Another critical point is breaking the chain whenever anomalies among our transaction patterns exceed the defined threshold.
	
	\subsubsection{Clue of Beginning Transaction of the Chain from Deposit Transaction}
\begin{figure}
		\centering
		\includegraphics[width=0.95\linewidth]{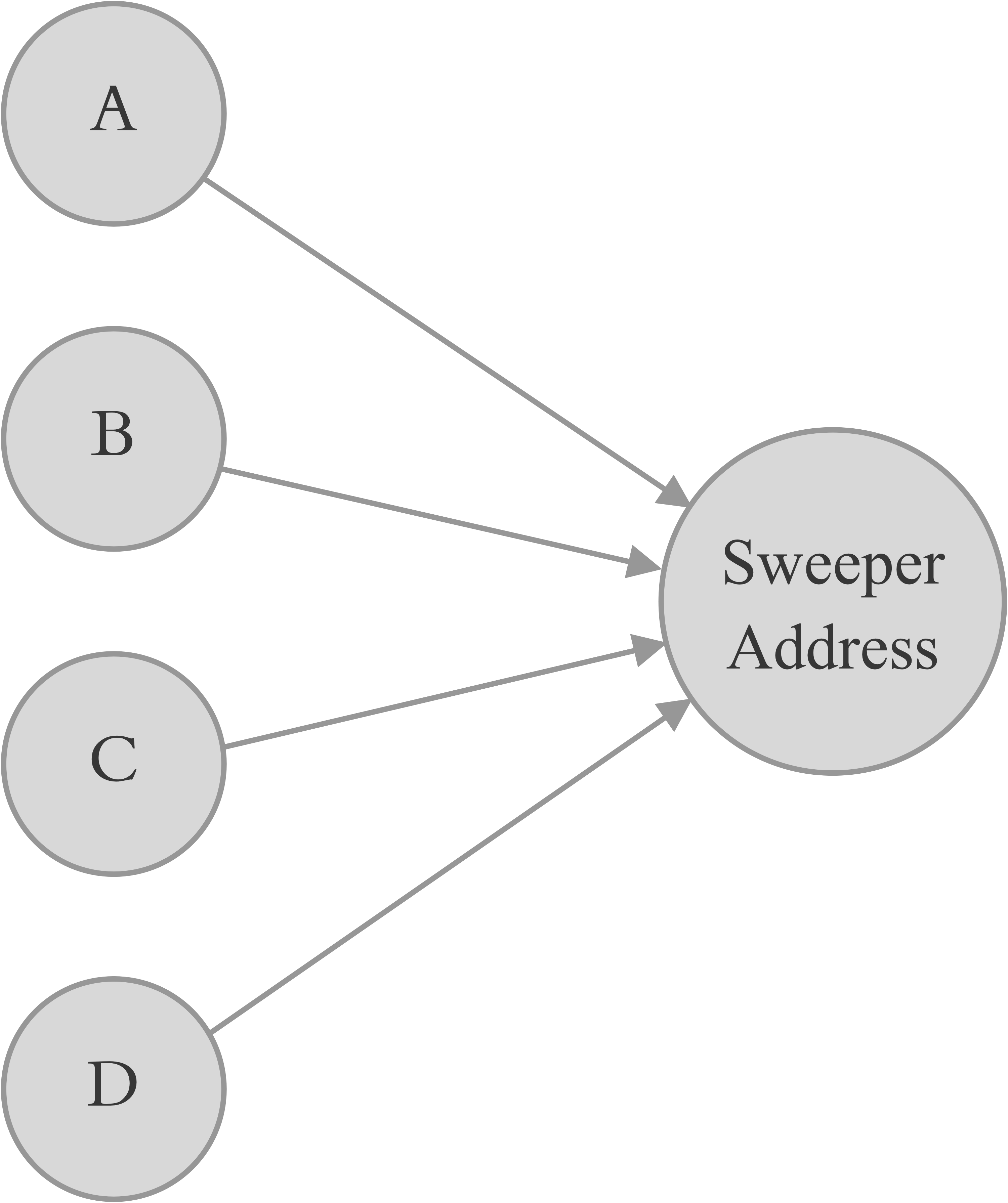}
		\caption{Sample real sweeper transaction beginning of transaction chain}
		\label{fig:sweeperTxBeg}
	\end{figure}

	Each time we send money to the mixing service, this money is usually not spent and stays at the mixing service address for a long time unspent. However, this unspent Bitcoin finally ended after a while. This period usually varied for the mixing service and the amount deposited at that address. Nevertheless, one issue was fixed between these mixing services; Finally, when this money (our deposit money to the service) was spent, the address was seen as input of a sweeper transaction. As we guessed, these transactions were aggregated to produce an address with a high amount of Bitcoin for a mixing service. As can be seen, one of these mixing sweeper transactions in Figure \ref{fig:sweeperTxBeg} shows the collection of money at various addresses belonging to the mixing service, which transfers them to an address with the help of a sweeper transaction for easier access and use in the intermediate stage of the transaction chain.
	
	\subsubsection{Investigating the Transaction Chain Started from Sweeper Transaction}
	As we mentioned in the previous section, The transaction of our money sent to the mixing service led us to a sweeper transaction whose inputs are addresses belonging to the mixing service with smaller amounts that aggregated the amounts in one address in the output. Then from that sweeper transaction, we form the forward chain of the transaction. After the formation of this transaction chain, as expected, the transactions after the sweeper transaction were transactions with the pattern of mixing transactions that we introduced in algorithm \ref{alg1}. The resulting chain was a chain of mixing transactions toward clean money (delivery of dirty money to the destination in a clear and disconnected path). Finding such a chain reduces the speed of finding transactions and the time complexity and increases the recall of our algorithm. If we wanted to act alone with the pattern obtained from the transaction, we had to randomly search and examine each of the transactions within the blockchain to reach mixing transactions. However, using the transaction chain, it is enough to find one of the transactions as a suspected candidate transaction of mixing service then expand the chain from it both forward and backward with the pattern of the transaction chain and the formation of this chain in the form presented in algorithms \ref{alg1} and \ref{alg2} and quickly find other mixing transactions as well.

	The problem that remains unresolved despite the formation of the transaction chain is the problem of false positives. We have not yet applied a new filter to reduce false positives. Next, we complete the two-phase filter method and filter more transactions with the help of a single-layer transaction chain.
	
	\subsubsection{Transaction Chain Length Pattern}
	After statistically examining the structure of transaction chains of mixing services, we concluded that the median length of these chains is 52, which is less than 10 for the same patterns in wallets. This heuristic can effectively reduce false positives related to wallets, but it is not suitable for exchanges with the same pattern as we considered for the transaction chain of mixing services.
	
	\subsubsection{Transaction Chain Time Pattern}
	By further investigating the transaction chains of mixing services statistically, we finally came to a pattern that reduced the false positives associated with exchange services. The average interval between transactions related to the mixer service included in the transaction chain is about 32 minutes, which is more than 10 minutes in transactions related to exchange services and more than 100 hours in transactions related to the wallets. This heuristic can be used for detecting and excluding a reasonable portion of transaction chains unrelated to mixing services.
	
	\subsection{Extracting Mixing Service Addresses in the Network}
	Based on the analysis we did in the previous sections, we were able to find and tag some transactions related to mixing services or money laundering. According to the analysis, we also came up with patterns for addresses, which we re-introduce below:
	\begin{enumerate}
		\item In intermediate transactions of extracted transaction chains; The entry address belongs to the mixing service.
		\item In intermediate transactions extracted transaction chains; The output address, which is P2SH and has a higher value, belongs to the mixing service (if both addresses were P2SH, we would only label them according to the amount).
		\item In the transaction start of the extracted transaction chains (sweeper transaction); All addresses belong to the mixing service.
	\end{enumerate}
	
	\subsection{Extracting Addresses Related to User(s) of Mixing Services}
	We tagged the addresses of the money-laundering service in the transactions detected. In this step, we want to find and tag the addresses of money launderers(mixing service users).
	
	\subsubsection{Clean-Side Money Addresses}
	The intermediate transactions of the extracted transaction chains are related to sending money laundered from the mixing service to the destination address. According to all the patterns and analyses that we have presented, we can say that the addresses related to the destination of the money launderer or money launderers are the address that has a lower amount in the output of the intermediate transactions of the transaction chains extracted from the algorithm \ref{alg3} which is not P2SH. Of course, if both addresses are P2SH, we only use the amount as a criterion. As mentioned earlier, the addresses obtained in this method are the addresses of money launderers on the clean side of the cycle (end of the money journey!).
	
	\subsubsection{Dirty-Side Money Addresses}
	Another address that can be extracted from mixing service users is the address of dirty money(incoming money), which can be accessed by tracking back the transaction entries of the sweeper transaction of the transaction chain(first transaction of the chain) related to mixing services. The algorithm for labeling the addresses of money launderers(mixing service users) shows the dirty money side in Algorithm \ref{alg4}.
	
	\begin{algorithm}[H]
		\caption {Identifying dirty addresses of mixing service user} \label{alg4}
		\textbf{Input}:
		A Sweeper Transaction\\
		\textbf{Output}: Address of the Sender (Dirty Money Addresses)
		\begin{algorithmic}[1]
			\WHILE{Tx Has Unchecked Input}
			\STATE address $\leftarrow$ pop address from input of tx
			\STATE transaction $\leftarrow$ previous transaction of address(redeemed transaction)
			\STATE \textbf{Label input addresses of transaction as Dirty Addresses}
			\ENDWHILE
		\end{algorithmic}
	\end{algorithm}
	
\begin{figure}
		\centering
		\includegraphics[width=\linewidth]{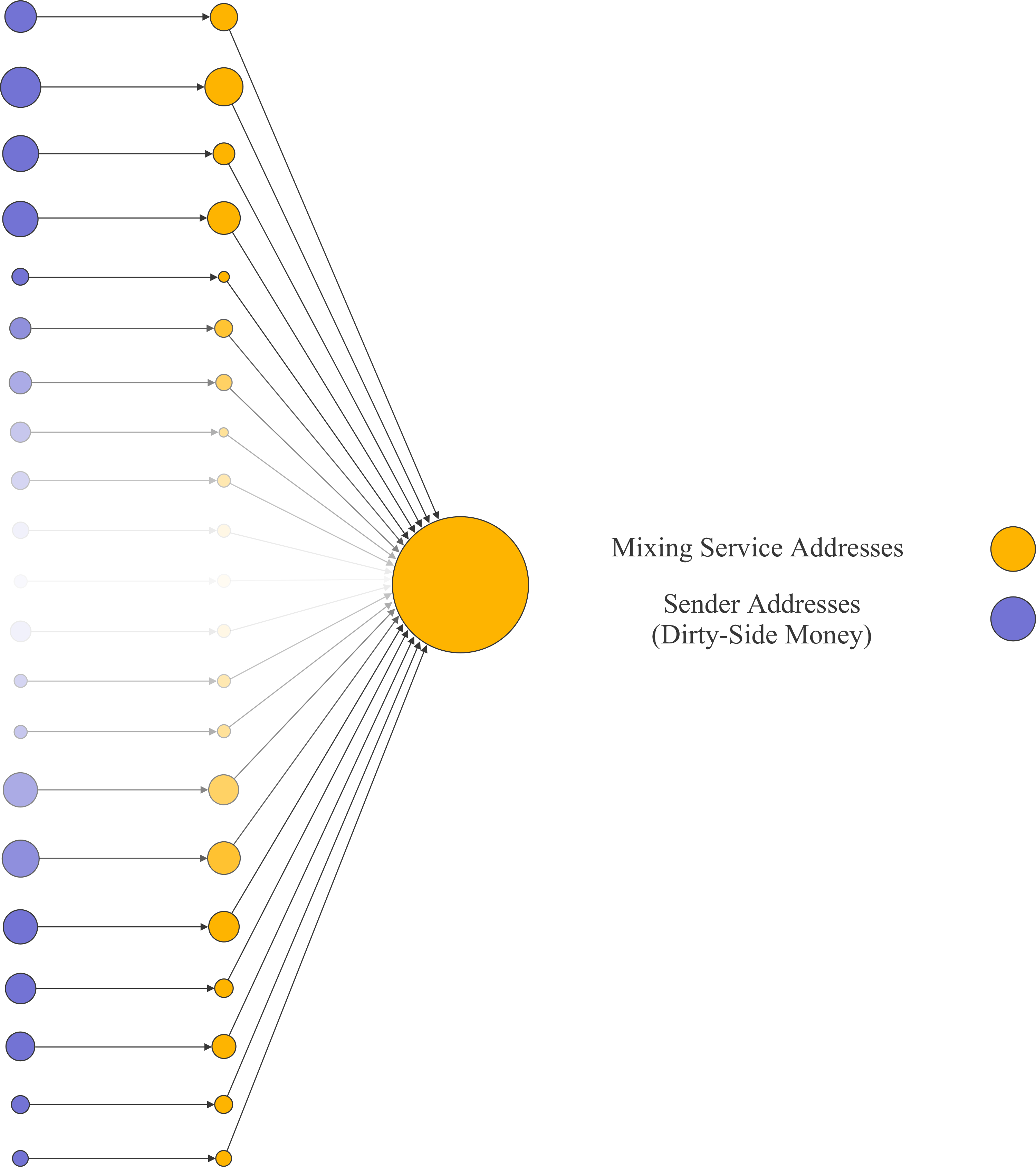}
		\caption{Dirty-Side Money and Connection of It to the Transaction Chain}
		\label{fig:dirtySide}
	\end{figure}

	In figure \ref{fig:dirtySide} we indicate the mechanism of detecting the dirty side of the Bitcoin, which is deposited to the mixing service again in a schematic view for better understanding.
	
	\subsection{What are the possible causes of the obtained patterns?}
	
	With a closer look at the obtained patterns, we can guess the reasons for their existence.
	In the early days of bitcoin, if people wanted to own bitcoins jointly, they often used MultiSig addresses. However, since 2012, when P2SH addresses were added to the official Bitcoin client, Due to the smaller transaction and the reduction of transaction fees, and greater ease of operation, the popularity of these addresses increased for multi-owner assets instead of multi-signature addresses. Based on this information, the pattern of address types can be analyzed so that since mixing services probably have more than one owner, the addresses related to these services are often P2SH.

	Another pattern is the one-to-two structural pattern of intermediate transactions. As we have already explained the function of today's mixing services, they transfer the amount to the address or outgoing addresses from another address where they have much money and in a separate path from senders money (this is the reason for the amount pattern) and most of the time a single outgoing address is considered. Another address is the change address of the rest of the Bitcoin amount of mixing service.
	
	\subsection{Expanding Identified Addresses}
	As mentioned in the research background, most of the efforts have been made to identify the category of addresses and identify the addresses associated with an identity and group them, and we have used these efforts to increase our data.
	
	\subsubsection{Clustering Addresses}
	As we described previously, as the first attacks on bitcoin anonymity in research, they provided a way to group and cluster addresses, which in practice is the best and most accurate way to reveal at the level of address grouping. In this method, it is assumed that the addresses in an entry belong to a single person.

	We have also expanded the categorized addresses, both for money launderers and money laundering services, as far as possible.
	
	\subsection{Labeling Detected Clusters}
	The grouping of addresses and transactions is the primary purpose of this study. However, according to previous research to more accurately label and match real-world information to addresses within the Bitcoin network, we also used this research in our own way and more information. We have obtained addresses from some groups, which we will explain below.
	
	\subsubsection{Correspond IP Addresses to Network Address Groups}
	Another method previously discussed in the previous section is to personalize the Bitcoin client to store the IP address of the transaction sending node on the network. We implemented this way with a more secure architecture. We personalized the client and implemented these clients on five servers in different parts of the world, namely Iran, China, Australia, Russia, and the United States, and finally labeled address groups by IP addresses we gathered.

	Of course, this method is wrong in the case of using TOR and the same services. However, if the IP is the same for different addresses within an address cluster, it gives us more reliability for the result.
	
	\section{Two-Phase Detection Algorithm}
		Based on the recurring patterns we discovered at the transaction and chain levels in the previous section, we propose a two-step algorithm for identifying mixing transactions. In this algorithm, the transactions are checked at the transaction level, and suspicious transactions are identified in the first step. In the second step, suspicious transactions are examined in more detail at the chain level and, if the conditions are met, are considered as mixing transactions. The addresses of the mixing service and its users are also determined according to the intended heuristics.
			
		\subsection{Implementing Identifier of Suspected Transactions}
		In this section, we implement an algorithm to identify suspected transactions related to withdrawal transaction in mixing services according to the obtained patterns.
		\begin{algorithm}[H]
			\caption{Finding suspected transactions of a block as a candidate mixing transaction} \label{alg1}
			\textbf{Input}:
			A Block of Transactions\\
			\textbf{Output}: Intermediate Mixing Transactions List of the Block
			\begin{algorithmic}[1]
				\REPEAT
				\STATE Ensure transaction has exactly 1 input 
				\STATE Ensure transaction has exactly 2 outputs 
				\STATE Ensure address types XOR equals 1
				\STATE Ensure one of outputs is 5 times or more greater than another one
				\STATE Ensure value of input is more than 1 BTC 
				\STATE If all above conditions met \textbf{Then} $\mathbf{CandidateMixingTranactions[]}\leftarrow\mathbf{transaction}$ 
				\UNTIL{All transactions have been checked}
			\end{algorithmic}
		\end{algorithm}
		Algorithm \ref{alg1} represents this implementation, although this algorithm alone is not very accurate, and we will talk more about it and its expansion later.
		
		\subsection{Implementation of a Two-Phase Algorithm for Detecting Mixing Transactions with High Accuracy}
		This section seeks to integrate the patterns obtained in an algorithm and reach the final method of identifying mixing transactions with high accuracy.
		In the algorithm \ref{alg1} function, we identified at the transaction level, and in the algorithm \ref{alg2} function, we used the first algorithm and achieved a higher level of abstraction by introducing the transaction chain to other transactions. In this section, we will integrate these algorithms into a general finalized algorithm with enhanced accuracy and recall due to both two-phase filtration and chain expansion.
		
		\begin{algorithm}[H]
			\caption{Final two-phase algorithm} \label{alg3}
			\textbf{Input}:
			A Block As a Seed of Algorithm\\
			\textbf{Output}: Labled Mixing Addresses , Labled Dirty Money Addresses, Labled Cleaned Money Addresses
			\begin{algorithmic}[1]
				\STATE {onceFilteredTxs} $\leftarrow$ {Algorithm1(Input Block)}
				\WHILE{Exist onceFilteredTxs}
				\STATE Potential Tx $\leftarrow$ Pop Tx
				\STATE {transactionChain , anomaliesDetected} $\leftarrow$ {Algorithm2(Potential Tx)}
				\IF{ anomaliesDetected > 5 } \STATE Continue \ENDIF
				\IF{Length(transactionChain) < 53 } \STATE Continue \ENDIF
				\STATE {timeMedian} $\leftarrow$ {MEDIAN( timeDifference(transactionChain) )}
				\IF{ timeMedian > 80 OR  timeMedian < 20  } \STATE Continue \ENDIF
				\STATE \textbf{Append transactions of chain to the database}
				\STATE \textbf{Label input addresses as Mixing Service Address}
				\STATE \textbf{Label greater P2SH output address as Mixing Service Address}
				\STATE \textbf{Label non-P2SH output address as clean recipient address(Destination)}
				\STATE \textbf{Algorithm4 ( transactionChain [0] ) }  
				\ENDWHILE
			\end{algorithmic}
		\end{algorithm}
		
		Algorithm \ref{alg3} receives a block as input and returns mixing transactions with high confidence. Note that in Algorithms \ref{alg2} and \ref{alg3}, several thresholds were set for exceptions in the pattern. For example, relay transactions are rarely seen among intermediate chain transactions that do not fit into our pattern. In this method, all patterns are used, and false positives are no longer high, while the execution time speed is much more efficient than having only Algorithm \ref{alg1}. 
		
\begin{figure}
			\centering
			\includegraphics[width=\linewidth]{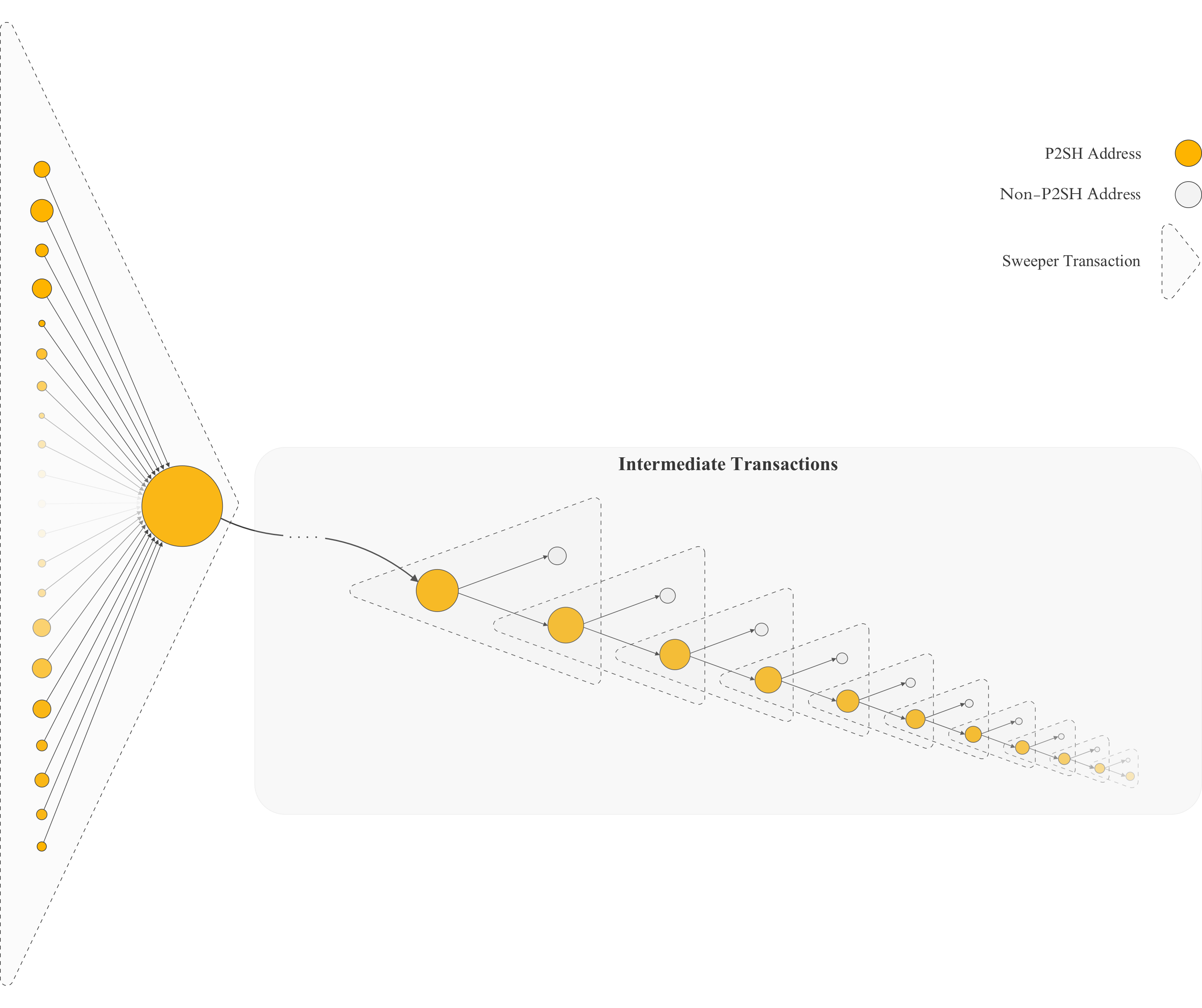}
			\caption{Full chain of mixing transactions concluded by third algorithm}
			\label{fig:fullChain}
		\end{figure}

		In figure \ref{fig:fullChain}, we provided a schematic of the final result of algorithm \ref{alg3} as all mixing service transactions we detect.
		
	\section{Conclusion}
	In this paper, we proposed an algorithm for detecting mixing transactions based on the statistical patterns we observed in the mixing process. Due to the lack of reliable labeled data, we selected three well-known mixing services MixTum, Blender, and CryptoMixer, and by using their service, we created mixing transactions. By investigating these transactions, we found recurring patterns at both transaction-level and chain-level.
	
	By examining all the transactions in the blockchain, matching them with the discovered patterns, and discovering mixing transactions, we distinguished between addresses related to mixing services and addresses related to the users of mixing services. The precision of our method is 100\%. However, we cannot compute recall and accuracy due to the cloudy atmosphere of the problem of detecting mixing services.
	\section{Future Works}
	
	In our proposed method, deposit and withdrawal transactions are not linked to each other, and only the addresses related to users of the service are separated from the addresses related to the service itself. Finding the link between the input and output money in the mixing process can be a significant step in tracking money after applying the mixing process.
	

\end{document}